\documentclass[aps,prl,twocolumn,superscriptaddress]{revtex4-1}
\usepackage{graphicx}   
\pdfoutput=1

\newcommand{\Eq}[1]{Eqn.~\ref{#1}}
\newcommand{\eqn}[1]{eqn.~\ref{#1}}

\newcommand{\Fig}[1]{Fig. \ref{#1}}

\usepackage{color}
\usepackage{hyperref}

\begin{document}

\title{Topological mechanics of gyroscopic metamaterials}

\author{Lisa M. Nash}
\affiliation{James Franck Institute and Department of Physics, The University of Chicago, Chicago, IL 60637, USA}

\author{Dustin Kleckner}
\affiliation{James Franck Institute and Department of Physics, The University of Chicago, Chicago, IL 60637, USA}

\author{Alismari Read}
\affiliation{James Franck Institute and Department of Physics, The University of Chicago, Chicago, IL 60637, USA}

\author{Vincenzo Vitelli}
\affiliation{
Instituut-Lorentz for Theoretical Physics, Universiteit Leiden, 2300 RA Leiden, The Netherlands}

\author{Ari M. Turner}
\email[]{aturne26@jhu.edu}
\affiliation{Department of Physics and Astronomy, Johns Hopkins University, Baltimore, MD 21218, USA}

\author{William T.M. Irvine}
\email[]{wtmirvine@uchicago.edu}
\affiliation{James Franck Institute and Department of Physics, The University of Chicago, Chicago, IL 60637, USA}

\begin{abstract} Topological mechanical metamaterials are artificial structures whose unusual properties are protected very much like their electronic and optical counterparts.
Here, we present an experimental and theoretical study of an active metamaterial -- comprised of coupled gyroscopes on a lattice -- that breaks time-reversal symmetry. 
The vibrational spectrum of these novel structures displays a sonic gap populated by topologically protected edge modes which propagate in only one direction and are unaffected by disorder.
We present a mathematical model that explains how the edge mode chirality can be switched via controlled distortions of the underlying lattice.  
This effect allows the direction of the edge current to be determined on demand.
We envision applications of these edges modes to the design of loss-free, one-way, acoustic waveguides and demonstrate this functionality in experiment.
\end{abstract}

\maketitle

\begin{figure*}
\includegraphics[]{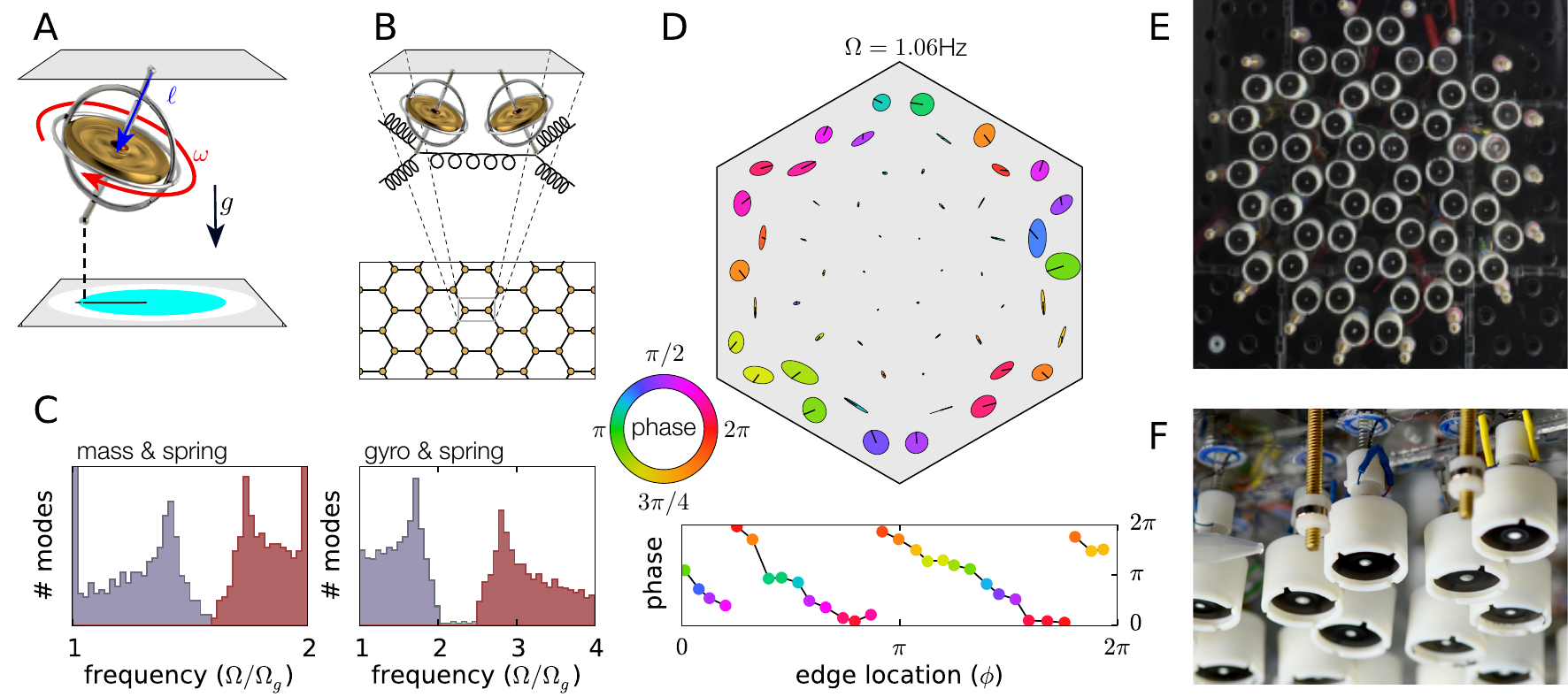}
\caption[]{
	Gyroscopic metamaterials. 
	\textit{(A)} The tip of a single gyroscope in a gravitational field precesses in a circular orbit.
 	\textit{(B)} A metamaterial can be constructed by connecting many gyroscopes with springs in a regular pattern, in this case a honeycomb lattice.
	\textit{(C)} A comparison between the density of states of a mass-spring and gyroscope-spring lattice.
	The forces on the gyroscope network includes a gravitational term; to allow direct comparison a central pinning force is included for the mass network. 
	(In both cases, $\Omega_g = \Omega_k = 1$.) 
	Because the honeycomb lattice is under-coordinated, the mass spring system has a large number of modes at $\Omega = 1, 2$.
	\textit{(D)} A chiral edge mode observed in an experimental metamaterial composed of 54 gyroscopes. 
	The shape of each orbit in the normal mode is indicated with ellipses (actual amplitude is $\sim1/10$ the indicated size, relative to the lattice spacing), and the phase at a fixed time is indicated via color.
	The phase of the gyroscopes along the edge is indicated below, showing the chirality of the mode.
	\textit{(E)} A picture of the experimental system viewed from below.
	\textit{(F)} The edge of the experimental lattice from the side, showing the construction of the individual gyroscopes.
	The air nozzle used to excite the gyroscopes is visible on the left.
	}
\label{fig:fig1}
\end{figure*}

A vast range of mechanical structures, including bridges, covalent glasses and conventional metamaterials (like origami, linkages and buckling metamaterials) can be ultimately modeled as networks of masses connected by springs~\cite{Maxwell1864, Wei2013, Thorpe1983, Wyart2005, Lubensky2015, Kang2014}.
Recent studies have revealed that despite its apparent simplicity, this minimal setup is sufficient to construct topologically protected mechanical states~\cite{Kane2013, Paulose2015, Chen2014, Paulose2015a, Vitelli2014} that mimic the properties of their quantum analogues~\cite{Hasan2010}.
Irrespective of its classical or quantum nature, a periodic material with a gapped spectrum of excitations can display topological behavior as a result of the non-trivial topology of its band structure~\cite{Prodan2009, Wang2009, Berg2011, Yang2015, HafeziM.2013, Rechtsman2013, Susstrunk2015, Peano2014, Jia2013}.

Most mechanical systems are invariant under time reversal because their dynamics are governed by Newton's second law, which, unlike Schr\"odinger equation, is second order in time. 
Theoretical work has suggested that if time-reversal symmetry is broken~\cite{Prodan2009},  as in  recently suggested acoustic structures containing circulating fluids~\cite{Yang2015}, phononic chiral topological edge states which act as unidirectional loss-free waveguides could be supported.
In this letter, we show that by creating a coupled system of gyroscopes and springs, a `gyroscopic metamaterial', we can produce an effective material with intrinsic time-reversal breaking.
As a result, our gyroscopic metamaterials support topological mechanical modes analogous to quantum Hall systems, which have robust chiral edge states~\cite{Haldane2008, Raghu2008, Wang2008}.
We demonstrate these effects by building a real system of gyroscopes coupled in a honeycomb lattice.
Our experiments show long-lived, unidirectional transport along the edge, even in the presence of significant defects.
Moreover, our theoretical analysis indicates that direction of edge propagation is controlled both by the gyroscope spin and the geometry of the underlying lattice.
As a result, deforming the lattice of gyroscopes allows one to control the edge mode direction, offering unique opportunities for engineering novel materials.

Much of the counterintuitive behavior of rapidly spinning objects originates from their large angular momentum, which resists changes in the axis of spinning.
If we fix one end of a gyroscope and apply a force, $\vec F$, to the opposing free end of the spinning axis, we produce a torque of $\vec{\tau} = \vec r \times \vec F$, where $\vec r$ is the axis of the gyroscope, pointing from the fixed to the free end.
In the fast spinning limit, the response of the gyroscope's axis is:
\begin{equation}
\label{eqn-gyro}
\frac{d\hat{r}}{dt} = \frac{\vec{r}}{I \omega}  \times \vec{F},
\end{equation}
where $\omega$ is the gyroscope angular frequency and $I$ is its rotational inertia. 
The behavior of a gyroscope differs from that of a simple mass in two important ways: (i) it moves perpendicular to applied forces and (ii) its response is first order in time.
The canonical example of this unusual behavior is precession: a spinning top does not simply fall over, but rather its free end orbits around the contact point (precesses) with a constant period, $\Omega_g = \frac{m g \ell}{I \omega}$, where $\ell$ is the distance from the pivot point to center of gravity, (\Fig{fig:fig1}A).

What happens if we replace the masses in a conventional mechanical metamaterial with gyroscopes?
To illustrate the difference between these two systems, we perform a normal mode analysis of honeycomb lattices composed of mass-spring and gyroscope-spring networks.
The density of states of these two systems  (\Fig{fig:fig1}C) show qualitatively similar features: each is characterized by two bands: a lower `acoustic' band (where neighboring sites move in phase) and an upper `optical' band (where neighboring sites move out of phase). 
The connections between these two bands, however, show key differences: in the mass-spring system the two bands touch at a Dirac point, while in the gyroscopic system a gap opens up between the bands.
Crucially, this gap is not empty, but populated by nearly equally spaced modes; the number of these edge modes scales with the length of the edge.
Examination of these gap modes reveals them to be confined to the edge and to be chiral: the phases always rotate in the same direction as one moves around the lattice (\Fig{fig:fig1}D).
As we show below, these edge modes are topologically robust and can therefore serve as loss-free, unidirectional waveguides. 

One key characteristic of topological states is that they are resistant to disorder, suggesting that they may be useful for acoustic applications and observable under a wide range of experimental conditions. 
However, the correspondence to a simplified model along the lines of \eqn{eqn-gyro} is far from guaranteed: real systems will have mixed-order dynamics (e.g.~nutation), lattice imperfections, gyroscopic non-uniformity, etc.
To explore the relevance of these effects, we have assembled a prototype system of 54 interacting gyroscopes on a honeycomb lattice, \Fig{fig:fig1}E.
Our gyroscopes consist of small DC motors spinning cylindrical masses at  $\sim$300 Hz; each is suspended from a top plate by a weak spring, producing an individual precession frequency of $\Omega_g \sim 1$ Hz.
To couple these gyroscopes in a lattice, a small neodymium magnet is placed in each spinning mass, with its moment aligned vertically.
This produces a short-range repulsion between the gyroscopes, creating an effective spring force between them which is comparable to the gravitational force.
(See supplemental materials for experimental details.)

\begin{figure}
\includegraphics[]{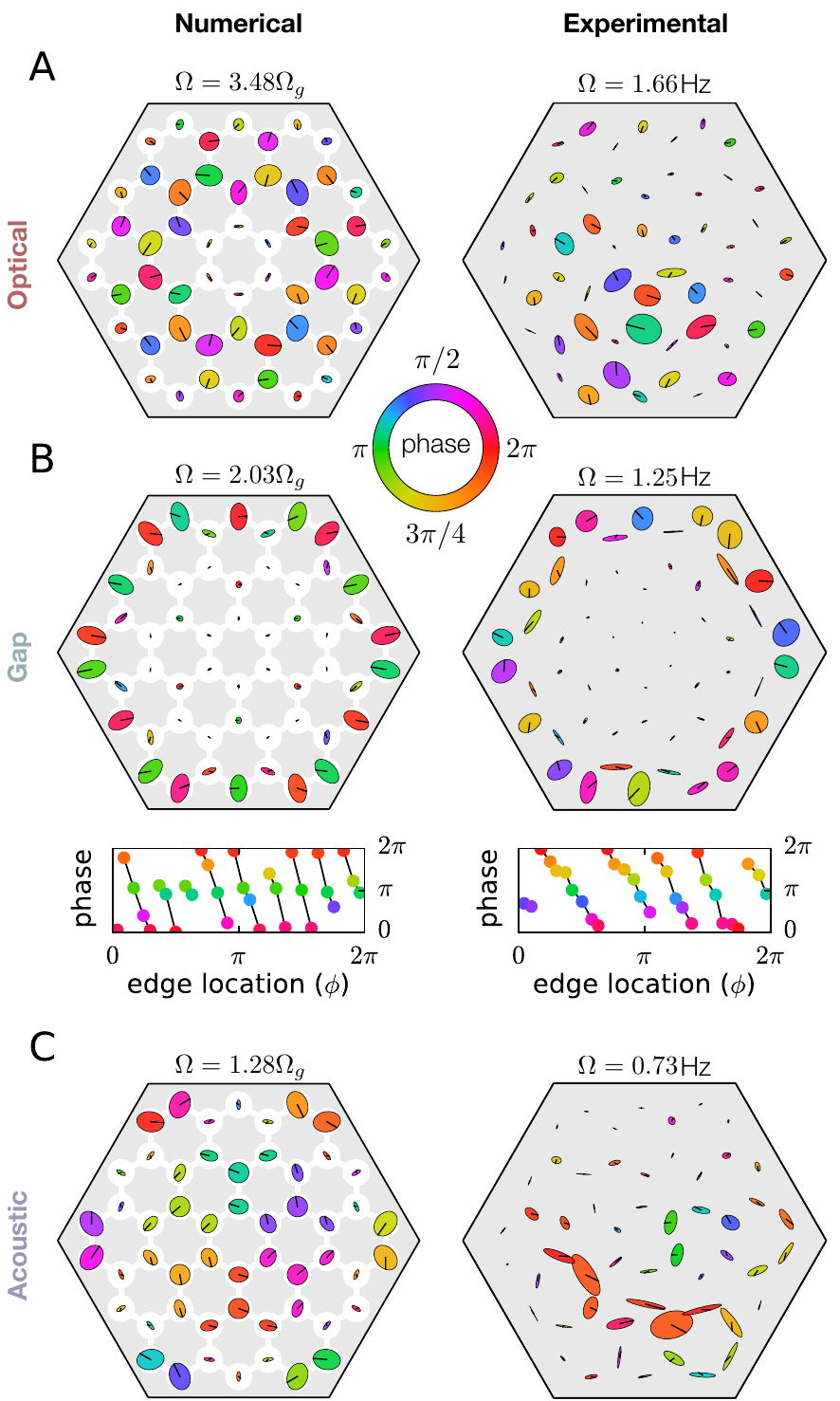}
\caption[]{
	A comparison of calculated normal modes in an ideal spring-gyroscope network (left) and as measured in an experimental system.
	For each system a mode is shown in \textit{(A)} the optical band, \textit{(B)} the bandgap, and \textit{(C)} the acoustic band.
}

\label{fig:fig3}
\end{figure}

\begin{figure*}
\includegraphics[]{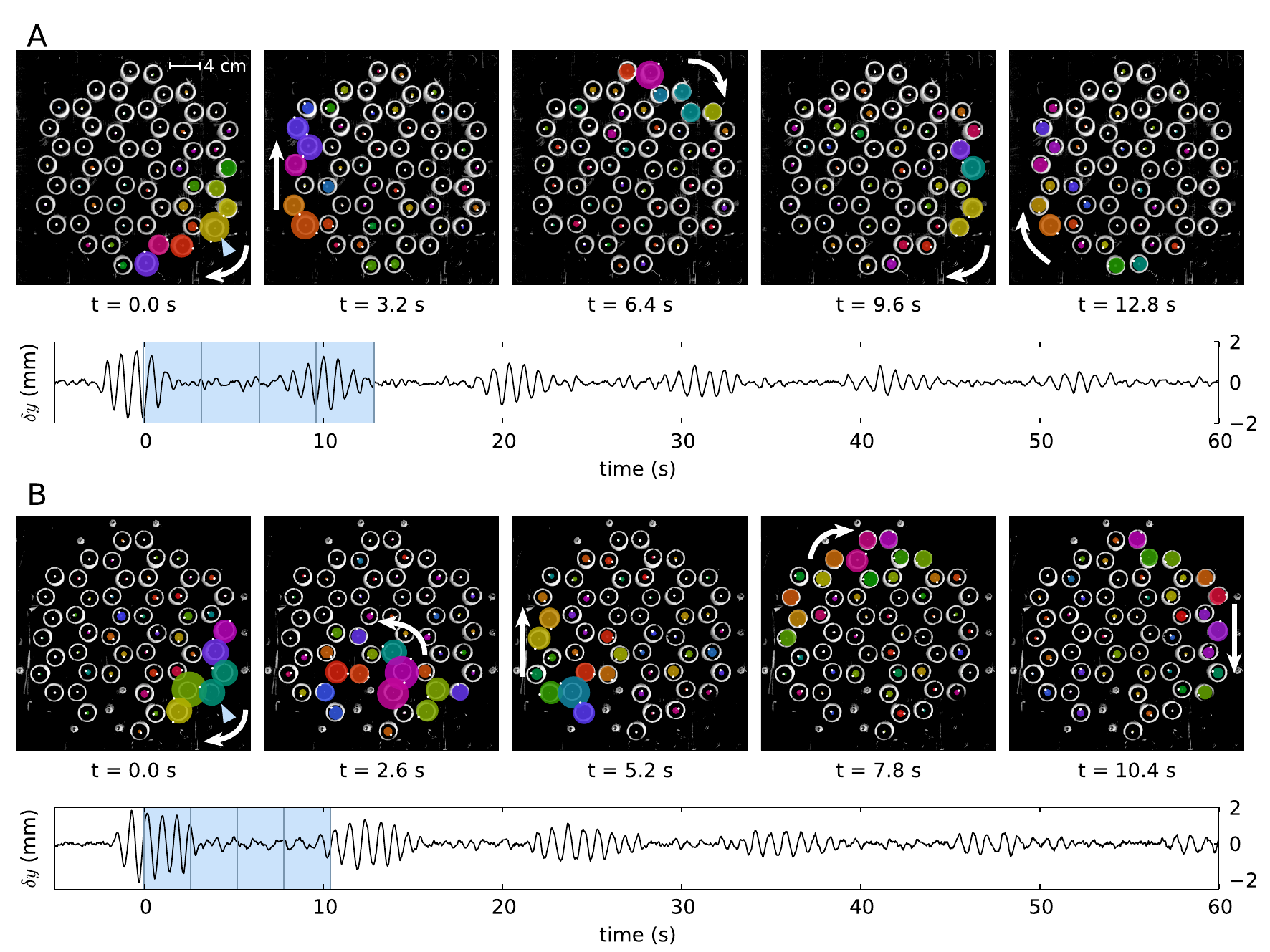}
\caption[]{
	Unidirectional waveguide modes in experiment.
	\textit{(A)} A single edge gyroscope is excited for five periods; subsequent images show the excitation propagation clockwise around the edge.
	The bottom graph indicates the displacement of one gyroscope (indicated with a triangle) in the $y$-direction; the excitation is seen to persist for many cycles around the edge.
	\textit{(B)} The same experiment as in \textit{(A)}, but with three gyroscopes removed from the bottom edge and replaced with fixed magnets (to keep the system in equilibrium).
	Due to the topological nature of the edge modes, the excitation propagates around the disturbance without losing amplitude or scattering.
}
\label{fig:fig4}
\end{figure*}

\begin{figure}
\includegraphics[]{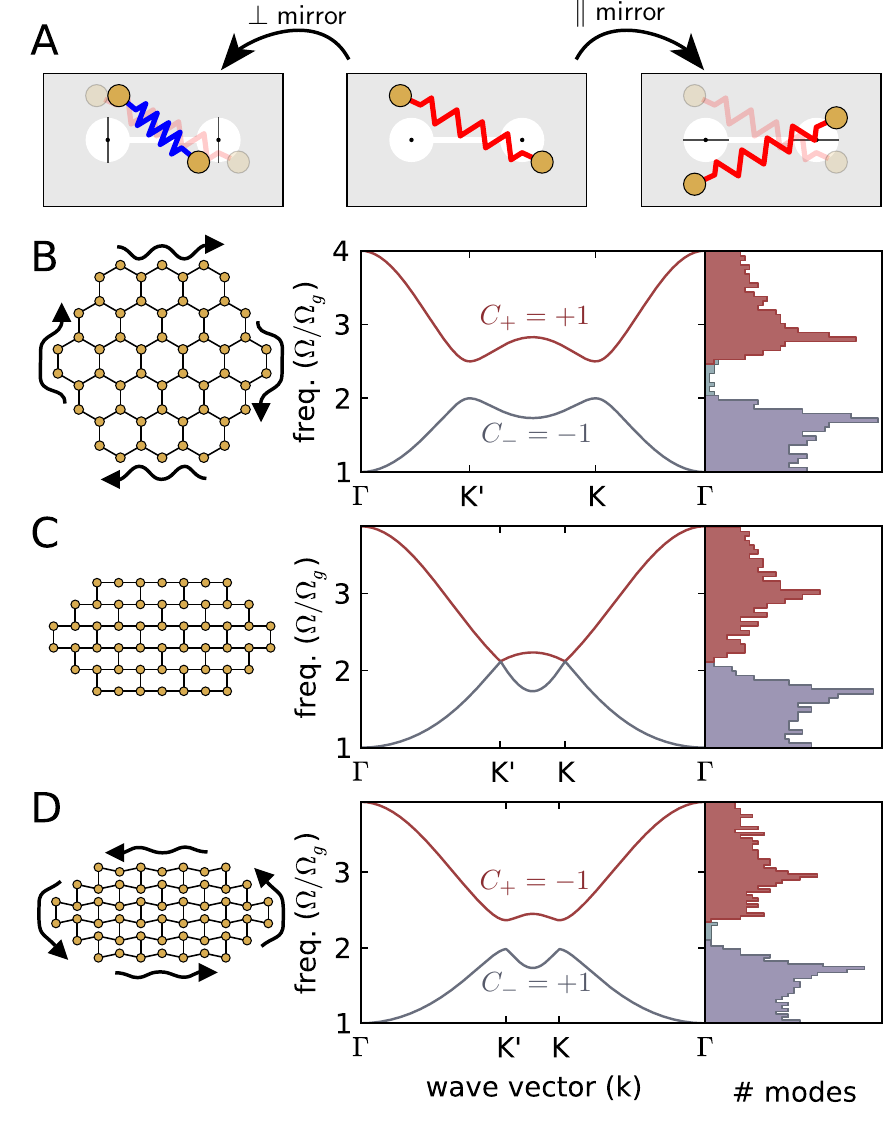}
\caption[]{
	Controlling time-reversal invariance in a gyroscopic metamaterial.
	\textit{(A)} The effect of mirroring a configuration of two gyroscopes around axes which are perpendicular or parallel to the equilibrium bond angle. 
Mirroring about axes that are perpendicular to the bond angle (left) converts stretching to compression but conserves the energy stored in the bond.  
Mirroring about the axes that are parallel to the bond (right) conserves the total bond length.  
Mirroring about other axes, in general, does not conserve energy.
	\textit{(B-D)} For each lattice geometry (shown at left), the band structure for an infinite system is shown (middle) along with the density of states for a finite lattice of 726 gyroscopes (right).
	\textit{(B)} shows an undisturbed, hexagonal honeycomb lattice.
	\textit{(C)} shows a honeycomb lattice as it is distorted into a rectangular configuration while preserving the connectivity, and \textit{(D)} shows a honeycomb lattice distorted past the square configuration.
	The rectangular configuration has no bandgap, and consequently no edge modes; as the lattice is further distorted the bandgap reopens but the edge modes have opposite chirality.
	}
\label{fig:fig2}
\end{figure}

To test the mechanical response of the gyroscopic metamaterial, we excite it with periodic bursts of air through a small nozzle, and follow the resulting disturbance.
To probe the normal modes, we weakly excite the gyroscopes at fixed frequencies for many ($> 100$) periods, and record the resulting motion of the network.
As shown in \Fig{fig:fig3}, experimental acoustic modes show in phase oscillation of adjacent gyroscopes (C) and experimental optical modes show out of phase oscillation of adjacent gyroscopes (A).
While the modes show qualitative agreement in phase of neighboring sites, 
there is little overall agreement in the mode structure.
This poor correspondence is due to the numerous non-idealities of the system: spread in the motor speeds ($\sim$10\%), error in gyroscope positions, next-nearest neighbor couplings, and only partially first-order dynamics (due to finite spinning speed).

Despite all of these imperfections, exciting in the gap between the acoustical and optical modes produces clean excitations along the edge (\Fig{fig:fig1}D $\&$ \Fig{fig:fig3}B).
The orientations and relative orbit sizes of these modes closely match the modes numerically computed for a simple idealized model.
To demonstrate that our experimental metamaterial functions as a unidirectional waveguide, we excite a single edge gyroscope for five periods at a frequency in the gap.
As shown in \Fig{fig:fig4}A and supplementary movie 1, the resulting excitation propagates in only one direction around the edge of the lattice.
The motion of this wave packet around the edge is persistent, circumnavigating the boundary several times.
As expected, short excitations at a frequency not in the band gap do not produce a similar robust edge excitation (supplemental movie 2).
Crucially, this indicates that the chiral edge modes are topologically protected from coupling to the bulk modes, functioning as an efficient one-directional waveguide.

We further demonstrate the robustness of these edge modes by intentionally introducing disorder in the lattice, for example, by removing three gyroscopes.
As shown in supplementary movie 3 and \Fig{fig:fig4}B, even this significant disturbance does not destroy the chiral edge modes.
An excitation on the edge is seen to move around this disturbance -- in the same direction as before -- and emerge undisturbed on the other side.
Remarkably, the excitation traverses the defect region without scattering backwards or into the bulk.
As before, the resilience of the edge modes indicates the topological character of the edge states in the gyroscopic metamaterial.

How can we understand the origin of this effect in our system? 
In general, symmetries play a fundamental role in characterizing a system's topological behavior; in the case of the gyroscopic materials broken time-reversal symmetry is the essential component.

To analyze the origin of these effects, we return to an ideal gyroscope-spring model.
For simplicity, we replace our vector gyroscopic axis, $\vec r$, with a complex representation of the tip displacement from equilibrium: $\psi \rightarrow r_x + i r_y$.
In this form, the linearized version of \eqn{eqn-gyro} is $i \frac{d \psi}{dt} = \frac{\ell^2}{I \omega} F$, where $F \rightarrow F_x + i F_y$ is the complex representation of the force and the complex phase, $i$, arises from the cross-product.
Accordingly, the linearized equation of motion for each site in the gyroscopic metamaterial is:
\begin{equation}
	\label{lattice-eom}
	i \frac{d \psi_p}{d t} = \Omega_g \psi_p +\frac{ \Omega_k}{2} \sum_q^{n.n.}  \left[\left(\psi_p - \psi_q\right) +  e^{2 i \theta_{pq}}\left(\psi^*_p - \psi^*_q\right)\right],
\end{equation}
where $p$ is the site label, $q$ are the neighboring sites connected by springs, and $\theta_{pq} = \arctan \frac{y_q - y_p}{x_q - x_p}$ is the global spring bond-angle (where $x_p$ and $y_p$ are the coordinates of gyroscope $p$ in the lattice)~\footnote{We note that in the case of  gyroscopes interacting via a repulsive potential, such as the repulsive magnetic dipolar interaction used in our experiments, Equation~\ref{lattice-eom} is modified by the appearance of real unequal real coefficients multiplying the terms $(\psi^*_p - \psi^*_q)$ and $(\psi_p - \psi_q)$.  This does not affect the time reversal analysis.}.

We note that the linearized equation of motion bears remarkable similarity to the Schr\"odinger equation for the wavefunction of an electron in a tight-binding model. 
Thus, by analogy, we may analyze the breaking of temporal symmetry using the `time-reversal' operation in quantum mechanics: $t \rightarrow -t$, $\psi \rightarrow \psi^*$.
For gyroscopes, conjugating $\psi$ mirrors their displacement in the $y$-axis; applying the complete time-reversal operation to a single gyroscope leaves the equation of motion unchanged.
Similarly, for a network of gyroscopes \Eq{lattice-eom} is invariant under this operation only if the coefficient $e^{2 i \theta_{pq}}$ is real (up to a global rotation), and breaks the symmetry otherwise.
Thus, crucially, we see that the breaking of time-reversal symmetry depends on distribution of bond angles in the lattice, and not simply the response of individual gyroscopes.

The geometric origin of the time-reversal symmetry breaking can also be seen by considering the energy of a single spring connecting two gyroscopes.
In the linearized limit, the stretching/compression of the spring is given by: \hbox{$\Delta \propto \left(r_{x1} - r_{x2}\right) \cos \theta_{12} + \left(r_{y1} - r_{y2}\right) \sin \theta_{12}$.}
If we mirror each gyroscope in the $y$-axis ($\psi \rightarrow \psi^*$ or $r_y \rightarrow -r_y$), in general the spring length will be unchanged only if $\sin \theta_{12} = 0$, i.e.~if the mirror axis aligns with the equilibrium bond angle.
However, if $\cos \theta_{12} = 0$, or the mirroring axis is perpendicular to the bond, then the spring energy, $E_k \propto \Delta^2$, is conserved by converting stretching to compression (\Fig{fig:fig2}A).
When considering an entire lattice, we see that for arbitrary displacements the bond energy will be conserved under time-reversal if (and only if) we are able to choose a global mirror axis to which all bonds are either perpendicular or parallel.
As a result, time-reversal invariance is only guaranteed for lattices composed of square or rectangular building blocks.

As implied by the form of \eqn{lattice-eom}, our gyroscopic metamaterial has a  quantum-mechanical analogue: in the limit that the spring coupling is weak, $\Omega_k = \frac{k \ell^2}{I \omega} \ll \Omega_g$, our system maps on to the Haldane model of an electronic system in a honeycomb lattice (see supplemental materials for details)~\cite{Haldane1988}.
In the Haldane model, time reversal symmetry is broken by a staggered magnetic field.  
This field can be varied, resulting in a change in the topological character of the modes as quantified by the Chern number~\cite{Avron1983,Thouless1982, Hasan2010}.
Accordingly, depending on the strength of the field and asymmetry between the two sites in the unit cell, the Chern number of the bottom band is $C_- = 0, \pm 1$, and $C_+ = -C_-$ in the top band.
A Chern number of zero indicates a trivial topology (a normal insulator) while a non-zero Chern number indicates a non-trivial topology.
Whenever $C_\pm \neq 0$, topological edge modes appear in the gap between the two bands; the chirality and direction of propagation of these modes depends on the sign of the Chern number for lower band.

In gyroscopic metamaterials, the analogue to changing the magnetic field is to geometrically distort the lattice.
In either case, the relevant operation produces a phase-shift in the hopping between neighboring sites; in the gyroscope system this phase shift is determined by the bond angles, $\theta_{pq}$. 
In the case of an undistorted honeycomb lattice, the modes have a Chern number of $C_\pm = \pm 1$, in agreement with the Haldane model.
In a honeycomb lattice, it is possible to distort the constituent hexagons without changing the bond length (\Fig{fig:fig2}B-D), allowing us to change the  gyroscopic phase between neighboring sites without changing the network connectivity.
As predicted by the time-reversal analysis above, the band-gap and chiral edge modes disappear when the bonds fall on a rectangular grid (in which case $e^{2 i \theta_{pq}} = \pm 1$).
Furthermore, the dispersion relationship of an infinite gyroscopic lattice in this configuration has Dirac points at the corners of the Brillouin zone (\Fig{fig:fig2}C); this is topologically equivalent to the dispersion relationship of a honeycomb network of springs and masses.
Continuing to distort the lattice past this point restores the band-gap, but the edge modes now have \emph{opposite} chirality, as reflected in an inversion of the bands and hence of the Chern number; $C_\pm = \mp 1$.
As a result, excitations along the edge now move in the other direction, opposite to the precession of individual gyroscopes. 
These effects can all be seen in supplementary video 4 that shows the simulated dynamics of an edge mode in a gyroscopic meta-material as it is being distorted.
Remarkably, this indicates the direction of the edge wave-guide can be controlled purely through geometric distortions of the lattice, analogous to an effect recently observed in 1D acoustic phononic crystals~\cite{Xiao2015}.

We have presented  an experimental proof of concept and theoretical analysis of a topologically protected unidirectional waveguide in a real mechanical metamaterial.
The origin of our topological edge modes is due to time-reversal symmetry breaking; our analysis indicates this arises from the combination of the chiral nature of the gyroscopes and the geometry of the underlying lattice.
Because the direction of the edge modes can be changed discontinuously with geometric distortions, in principle small displacements should be capable of inverting the edge mode direction.
This mechanism may have practical applications for creating direction-tunable materials, but it also suggests interesting non-linear effects should occur in the regime near the mechanically-induced topological phase transition.

The prototypical gyroscopic solids we have developed here are examples of active metamaterials: their design relies on the presence of internal motors that keep each gyroscope in a fast spinning state. 
An open challenge is to 
construct scalable gyroscopic metamaterials using nano-fabrication techniques (e.g.~MEMS) or active molecules that convert chemical energy into rotation very much like the motors powering each gyroscope~\cite{Prost2015, Palacci2013}. 
Such an implementation would pave the way towards realizing materials that support one-way, loss-less acoustic waveguides at a microscopic scale.

\bibliography{refs}

\end{document}